\documentclass[aip,amsmath,amssymb,onecolumn]{revtex4-1} 

\usepackage{xcolor}
\usepackage{mathtools}
\usepackage{comment}
\usepackage{caption}
\usepackage{subcaption}
\usepackage{placeins}

\usepackage{amsmath}
\usepackage{amsfonts}
\usepackage{mathrsfs}
\usepackage{chemformula}

\usepackage{makecell}

\usepackage{soul} 

\newcommand{\bs}{\boldsymbol}

\newcommand{\kb}{k_{\text{B}}}

\newcommand{\position}{\boldsymbol{x}}

\newcommand{\paramposition}{\tilde{x}}
\newcommand{\gradposition}{\boldsymbol{\nabla}_x}

\newcommand{\force}{\boldsymbol{F}}

\newcommand{\commentout}[1]{}

\newcommand{\particlenoise}{\boldsymbol{\mathcal{W}}}
\newcommand{\discreteparticlenoise}{\boldsymbol{W}}

\newcommand{\mobility}{\boldsymbol{\mathcal{M}}}

\newcommand{\chargedensity}{\varrho}

\newcommand{\ito}{ }

\newcommand{\Dlength}{\lambda_\text{D}}

\newcommand{\deleted}[1]{}

\newcommand{\MarginPar}[1]   
{\marginpar{\vskip-\baselineskip 
\raggedright\tiny\sffamily\hrule\smallskip{\color{red}#1}\par\smallskip\hrule}}

\newcommand{\MarginJBB}[1]   
{\marginpar{\vskip-\baselineskip 
\raggedright\tiny\sffamily\hrule\smallskip{\color{blue}#1}\par\smallskip\hrule}}
\usepackage[utf8]{inputenc}

\begin{document}

\title{Steric effects in induced-charge electro-osmosis for strong electric fields}

\author{J. Galen Wang}
\affiliation{Center for Computational Sciences and Engineering, LBNL}

\author{Daniel R. Ladiges}
\affiliation{Center for Computational Sciences and Engineering, LBNL}
\homepage{https://ccse.lbl.gov}

\author{Ishan Srivastava }
\affiliation{Center for Computational Sciences and Engineering, LBNL}

\author{Sean P. Carney}
\affiliation{Department of Mathematics, UCLA}

\author{Andy J. Nonaka}
\affiliation{Center for Computational Sciences and Engineering, LBNL}

\author{Alejandro L. Garcia}
\affiliation{Department of Physics and Astronomy, SJSU}


\author{John B. Bell}
\affiliation{Center for Computational Sciences and Engineering, LBNL}

\date{\today}

\begin{abstract}
We study the role of steric effects on the induced-charge electro-osmosis (ICEO) phenomenon using a recently developed mesoscale fluid model. 
A hybrid Eulerian-Lagrangian method is used to simulate the dynamics of discrete immersed ions in a thermally fluctuating solvent near a metallic plate embedded in the dielectric interface. 
We observe that the characteristic velocity scales almost linearly with electric field when the generated $\zeta$-potentials exceed the order of the thermal voltage, as opposed to a quadratic scaling predicted by Helmholtz-Smoluchowski equation, although qualitative agreement with experiments and theories is obtained at low electric fields. Our simulations reveal that the steric effects play a crucial role at strong electric fields, which is observed from the aggregation of ions towards the center of the metal plate instead of at the edges, and the overcharging of co-ions to the surface charge near the electric double layer. A comparison to a continuum electrolyte model also highlights significant differences in charge distribution and flow field that are attributed to the steric repulsion between ions. 
\end{abstract}

\maketitle

\section{Introduction}\label{sec:intro}
Electrokinetic phenomena typically occur near a charged surface that is submerged in an electrolyte subject to an external electric field. The charged surface attracts a layer of counterions (that is, ions with opposite charge to the surface charge) that breaks electroneutrality. If an external electric field is applied, the Lorenz force on the ions induces flow.
Traditionally, if the charged surfaces are fixed the resulting flow is referred to as electro-osmosis. If the surfaces are suspended in the solution, such as charged colloids, the motion of those particles is called electro-phoresis. 
These processes are widely used in scientific and industrial applications, for example, in protein purification \cite{hames1998gel,janson2011protein,lee2022microfluidic}, fuel cells \cite{Andersson_2010,JahnkeETAL_2016,Arsalis2019,hu2020review,alkhadra2022electrochemical}, and ion separation processes such as water desalination \cite{grimm1998waterpurification,schoch2008transport,jiang2015electrodialysis,alkhadra2022electrochemical}. Induced-charge electro-osmosis (ICEO) \cite{squires2004induced,squires2009induced} is a special type of electro-osmotic phenomena where surface charges, and consequently an electric double layer consisting of counterions, are induced on a conductive or polarizable dielectric surface.
One of the unique features of induced-charge electro-osmosis is that it can generate persistent flow with an alternating-current (AC) field, because after some charging time, the same flow pattern is generated when the electric field is reversed. 
The advantages of induced-charge electrokinetic flows over conventional electro-osmosis have led to significant progress in many areas, particularly in microfluidic pumps, mixers and valves, and particle trapping \cite{harnett2008model,bazant2009towards,bazant2011icek,schnitzer2012induced,davidson2014chaotic,ren2015induced,khair2020breaking,manshadi2020induced,yao2020on,feng2020recent,hashemi2020asymmetric,li2022icek}.

Although ICEO has received significant attention that has led to many fruitful results, several challenges and practical difficulties remain. First, although experiments at low electric fields qualitatively match the predicted quadratic scaling \cite{bazant2009towards,squires2009induced} with respect to field strength, they typically measure significantly lower velocities than predicted by theory \cite{soni2007iceo,bazant2009towards}. 
The exact physical reason behind the reduced velocities is still not well understood. Second, induced-charge electrokinetic flow seems to be suppressed with increasing electrolyte concentration \cite{gangwal2008induced}. Previous experiments barely observe any flow signal for concentrations higher than 0.1~M \cite{bazant2009towards,brown2000pumping,green2000fluid,green2002fluid,studer2004acekpump,levitan2005experimental,ramos2005pumping,bown2006aceo,urbanski2006fast,urbanski2007effect,bazant2007electrolyte,soni2007iceo,storey2008steric}, a concentration typical of biological systems. Third, there are very few experimental studies at the high electric fields (tens to hundreds of times the thermal voltage $k_B T/e$) that are relevant for real-world applications.
A number of effects have been proposed in the literature that could theoretically explain the flow properties at high field strengths, such as surface conduction \cite{khair2008surprising,schnitzer2012induced}, viscoelectric effects \cite{lyklema1961interpretation,dukhin1993non,lyklema1994slip}, and Faradaic reactions \cite{olesen2006ac,ramos2007linear,ramos2016ac}.
However, how these factors change the flow quantitatively or even qualitatively is not yet clear. Consequently, prior studies were often conducted in the low-concentration and low-electric-field regime to avoid any ambiguity even though this regime is not applicable for many induced-charge electrokinetics applications. 

Simulations provide powerful tools for elucidating the physics of induced-charge electrokinetics beyond low concentrations and weak fields, because they can access a broad span of length and time scales under a wide range of electric fields. The key mechanism at play is the development of an electric double layer that breaks electroneutrality near the charged surface, creating a diffuse layer of counterions whose thickness is roughly one Debye length, $\lambda_D$ \cite{debye1923theory,robinson2012electrolyte,newman2012electrochemical}.
A number of numerical approaches have been proposed to model electrokinetic flows including continuum methods such as the Poisson-Nernst-Planck (PNP) equations \cite{newman2012electrochemical} and fluctuating hydrodynamics (FHD) \cite{Land2}, and particle-based methods such as molecular dynamics (MD) \cite{plimpton1995fast,thompson2022lammps}. 
The PNP equations are able to capture some of the basic behavior of electrokinetic flows. However, they are missing important mesoscale effects such as steric effects, which are crucial in electrokinetic flows but have not been integrated in ICEO modeling yet \cite{feng2020recent}. Furthermore, the PNP equations can require fine meshes and small time steps at high concentrations where the double layer, whose thickness scales inversely to the square-root of concentration, is extremely thin. The PNP equations also do not account for the fluctuating hydrodynamics of the continuum solvent, which is an important mesoscale effect arising from the thermal motion of the solvent molecules. Such effects of thermal fluctuations are incorporated in FHD, which has successfully simulated electrolyte solutions \cite{peraud2016low,peraud2017fluctuation} at moderate electrolyte concentrations.
However, FHD is often not suitable for modeling electrokinetic flows, particularly at high concentrations. In this regime, the resolution required to resolve the double layer becomes sufficiently fine that there are too few ions per computational cell to justify the underlying assumptions implicit in FHD models. 
In contrast, particle-based methods can model systems down to the molecular scale \cite{bedrov2019mdelectrolytes,yao2022mdbatteries}; however, they have drawbacks as well: MD using an explicit solvent suffers from high computational costs because of the large number of solvent molecules that have to be modeled.
In addition, it can be difficult to recover properties such as viscosity and permittivity in an explicit solvent model which are indirectly specified by the choice of intermolecular potential. On the other hand, in implicit-solvent MD far-field hydrodynamic interactions, which are important in electrolyte solutions and biological cells, are not included. 

To overcome the above limitations, considerable recent effort has been devoted to the development of mesoscale simulation techniques that combine continuum methods with particle methods to model discrete particles in a fluctuating, continuum solvent.
Examples of this type of approach include Brownian dynamics (BD) \cite{sierou2001asd,SE_Multiblob_SD}, the general geometry Ewald-like method (GGEM) \cite{hernandez2007fast,zhao2017parallel}, the stochastic Eulerian Lagrangian method (SELM) \cite{atzberger2007stochastic,atzberger2011}, the stochastic force coupling method (SFCM) \cite{maxey2001localized, lomholt2003force, keaveny2014fluctuating, delmotte2015simulating}, and the fluctuating immersed boundary method (FIB) \cite{delong2014brownian}. Among these techniques, BD has been the most widely used with established success in many applications \cite{jardat1999bdelectrolytes,yamaguchi2011bdelectrolyte,byun2020bddendrite}. It uses a Green's function representation to model hydrodynamic interactions of particles suspended in a solvent in the Stokes flow regime, but it is computationally expensive to fully resolve many-body hydrodynamics because it involves inversion of a dense mobility matrix. BD is also limited to simple geometries where analytical calculations of the mobility matrix are possible, such as those in Refs.~\citenum{jeffrey1984mobilityfunctions2spheres} and \citenum{kim2005microhydrodynamics}. FIB addresses the computational cost issue of BD by using immersed boundary kernels \cite{peskin2002} to spread particle-scale forces onto the fluid mesh, which is then solved using well established fluctuating hydrodynamics (FHD) techniques. As a result, this method avoids the expense of inverting a mobility matrix for particles, while maintaining a description for hydrodynamic radii of particles via the size of immersed boundary kernels, which accurately describes particle diffusion. We recently developed an extension of the FIB method, called the Discrete Ions Stochastic Continuum Overdamped Solvent (DISCOS) method \cite{ladiges2020discrete} that further reduces the computational cost. Specifically, DISCOS introduces a coarse-grained approach, where the particle positions are updated using the velocities obtained on a coarse grid for solving FHD with a ``dry diffusion" correction implemented to recover ion's total diffusivity. DISCOS also incorporates a particle-particle, particle-mesh (P3M) approach \cite{Hockney:1988} for efficiently computing electrostatic forces. 

The DISCOS method has been validated by comparison with MD and theory in several scenarios, including the ion-ion correlation function and conductivity in triply periodic domains \cite{ladiges2020discrete}, the equilibrium distribution of ions in a confined channel \cite{ladiges2022ekdiscos}, and several electrokinetic flows in doubly periodic domains \cite{ladiges2022ekdiscos}. In particular, for regular electro-osmotic flows DISCOS predicts realistic charge distribution and velocity profiles that are closer to MD results than PNP simulations, demonstrating the importance of steric effects in electrokinetic flows \cite{ladiges2022ekdiscos}. 
Preliminary simulations of induced-charge electro-osmosis using DISCOS in our previous paper \cite{ladiges2022ekdiscos} further illustrate the importance of including steric effects and thermal fluctuations. Our simulations predicted a charge distribution that is distinctly different from one predicted by the PNP equations, resulting in a peak velocity that is 50\% higher in the DISCOS simulation.

In this paper, we use DISCOS to investigate the complex electrokinetic flows arising from ICEO in more detail. A range of electric fields and high molar concentrations are considered, moving beyond the weak-field regime, where the induced $\zeta$-potential is small compared with thermal voltage, to elucidate new physics about ICEO. 
As in Ref.~\citenum{ladiges2022ekdiscos} the simulations are in a three-dimensional channel bounded by a wall in the $y$-direction and doubly periodic in the other two directions. The configuration is set to mimic the canonical experimental setup of ICEO \cite{soni2007iceo} where a metal strip is placed between dielectric materials. In solving for the electric field we impose homogeneous Dirichlet conditions to model the metallic plate and homogeneous Neumann conditions to model the dielectric part of the walls. Our main goal in this work is to investigate the electric-field-dependent velocity in ICEO, and gain insight into the important physics behind specific velocity scalings with respect to electric field. 

The remainder of this paper is organized as follows. We provide a brief summary of the DISCOS algorithm in Section~\ref{sec:methods}, focusing on the treatment on the boundary conditions for the electric potential. We present simulation results in Section~\ref{sec:results} and show both flow properties and charge distributions obtained from DISCOS and compare to the continuum hydrodynamic model (that is essentially the Poisson-Nernst-Planck (PNP) model \cite{peraud2016low}). We further present a velocity scaling over a wide range of electric field strengths, and compare our observation with proposed mechanisms from the literature. We conclude with a summary of our findings in Section~\ref{sec:conclusion}. The Appendix discusses the effect of dry diffusion and the determination of optimal simulation parameters.

\section{Methodology}\label{sec:methods}

\subsection{DISCOS: a brief summary}\label{subsec:discos}

DISCOS models the behavior of discrete ions in a solvent in the overdamped limit. The solvent is treated by mapping forces on the particles to a grid, solving the fluctuating Stokes equation on that grid and then interpolating the resulting velocities back onto the particles.  The details of the method are presented in our earlier papers \cite{ladiges2020discrete,ladiges2022ekdiscos} so here we only give a brief summary. 

The force acting on each particle includes contributions from electrostatic forces, short range interparticle forces, wall interaction forces, and forces due to a specified external field:
\begin{equation}
    \boldsymbol{F}_i = \boldsymbol{F}_i^\text{E} + \sum_{j} \boldsymbol{F}_{ij}^\text{R} + \sum_{k} \boldsymbol{F}_{ik}^\text{W} + \boldsymbol{F}_i^\text{ext}.
    \label{eqn:forces}
\end{equation}
Here, $\boldsymbol{F}_{ij}^\text{R}$ is the short range force between particles $i$ and $j$, and $\boldsymbol{F}_{ik}^\text{W}$ is the short ranged interaction between particle $i$ and wall $k$. In each case the force is calculated using a Lennard-Jones (LJ) \cite{jones1924determination1,jones1924determination2,frenkel2001understanding} type potential
\begin{align}
    U^{\rm sr}(\paramposition;\sigma,\xi) =
    \begin{cases}
        \widehat{U}^{\rm sr}(\paramposition;\sigma,\xi) - \widehat{U}^{\rm sr}(\paramposition_c;\sigma,\xi), \vspace{1mm}&0 < \paramposition < 2.5\sigma \\
        0, & 2.5\sigma \ge \paramposition
    \end{cases},\label{eq:potential}
\end{align}
where $\paramposition$ is the radial distance from a particle, $\sigma$ is the van der Waals diameter, and $\xi$ is the magnitude of the potential. For inter-particle forces we employ a 12-6 potential,
\begin{align}\label{eqn:LJ126}
    \widehat{U}^{\rm sr}(\paramposition;\sigma,\xi) = 4\xi \left(\left(\frac{\displaystyle\sigma}{\displaystyle\paramposition}\right)^{12} - \left(\frac{\displaystyle\sigma}{\displaystyle\paramposition}\right)^6\right).
\end{align}
For particle-wall interactions we use a 9-3 LJ \cite{abraham1977structure} potential
\begin{align}\label{eqn:LJ93}
    \widehat{U}^{\rm sr}(\paramposition;\sigma_{\textrm{wall}},\xi_{\textrm{wall}}) = \frac{3^{3/2}}{2}\xi_{\textrm{wall}} \left(\left(\frac{\displaystyle\sigma_{\textrm{wall}}}{\displaystyle\paramposition}\right)^{9} - \left(\frac{\displaystyle\sigma_{\textrm{wall}}}{\displaystyle\paramposition}\right)^3\right).
\end{align}

The electrostatic forces $\boldsymbol{F}_i^E$ are computed using a particle-particle, particle mesh (P3M) approach \cite{Hockney:1988}.  We represent the charge on each particle using a Peskin kernel \cite{peskin2002,New6ptKernel} to map the charges to a charge density $\chargedensity$ on a grid. Using that charge density we solve Poisson's equation
\begin{equation}
  - \epsilon  \nabla^2 \phi = \chargedensity
    \label{eqn:poisson}
\end{equation}
on the grid to obtain the electric potential $\phi$.  We then compute the electric field $E = -\nabla \phi$, interpolate it to the particle locations and multiply by the charge of the particle to compute a provisional electrostatic force on the particle. This approach provides an accurate approximation of the interaction of particles sufficiently far apart.  For particles that are close to one another we replace the force computed from the grid with a direct Coulomb force calculation. The interaction of two nearby particles computed from Poisson's equation can be precomputed and tabulated as a function of their separation, as discussed in Ref. \citenum{ladiges2020discrete}.

Once all the forces on the particles have been evaluated, we can compute the velocities of the particles resulting from those forces in the overdamped limit.  We represent the particles using (possibly different) Peskin kernels and use that representation to map $\boldsymbol{F}_i$ onto a representation of the forces on a grid, denoted by $\boldsymbol{f}$.
We then solve the fluctuating hydrodynamics Stokes equations
\begin{align}
    \nabla p - \eta \nabla^2 \boldsymbol{v} &= \boldsymbol{f} + \sqrt{2 k_B T \eta}~ \nabla \cdot \mathcal{Z}     \label{eqn:stokes}\\
    \nabla \cdot \boldsymbol{v} &= 0,\nonumber
\end{align}
where $p$ is the pressure, $\eta$ is the fluid viscosity, $\boldsymbol{v}$ is the fluid velocity, $k_B$ is the Boltzmann constant, $T$ is the temperature, and $\mathcal{Z}$ is a symmetric Gaussian white noise tensor field. These equations are solved using a geometric multigrid preconditioned GMRES algorithm \cite{Cai2014} to obtain the velocity field on the hydrodynamic grid.  The resulting velocity field can then be interpolated onto the particles.

The process of computing velocities for the particles from forces on the particles using Eq.~(\ref{eqn:stokes}) implicitly defines a particle mobility operator, $\mobility$.
This mobility operator defines an overdamped Brownian dynamics for the particle motion
\begin{equation}
    \frac{d\position}{dt} =\mobility \force    + \kb T \gradposition \cdot \mobility   + \sqrt{2 \kb T} \mobility^{1/2}\ito \mathcal{Z}.
    \label{eqn:bd}
\end{equation}
The first and third terms correspond to the two terms on the right hand side of Eq.~(\ref{eqn:stokes}).  The second term is an It\^{o} correction that arises because $\mobility$ depends on the particle location.
This term is treated using the random finite difference version of Fixman's algorithm \cite{fixman1978simulation} introduced in Delong {\it et al.} \cite{delong2014brownian}. 
This approach adds a correction term to $\boldsymbol{f}$ in the Stokes equation coupled with a midpoint update for the particle locations.

In DISCOS, the choice of the grid spacing and Peskin kernel determine the effective hydrodynamic radius of a particle.  DISCOS incorporates a correction, referred to as ``dry diffusion'', that compensates for having a hydrodynamic radius that depends on the grid resolution.  This is important when modeling ions of different size and also enables using a coarser grid for the Stokes solve than would be dictated by the actual hydrodynamic radius of the particle.  With this correction the equation to advance the particles when dry diffusion is included is
\begin{align}
    \frac{d \position}{d t} = \underbracket[0.4pt][2pt]{\mobility \force    + \kb T \gradposition \cdot \mobility   + \sqrt{2 \kb T} \mobility^{1/2}\ito \mathcal{Z}
    \vphantom{ \frac{D^{\rm dry}}{\kb T}\force }       }_{\text{wet}} 
    + \underbracket[0.4pt][2pt]{ \frac{D^\text{dry}}{\kb T} \force  +  \nabla_x \cdot D^\text{dry}+ \sqrt{\frac{2 D^\text{dry}}{\Delta t}}\ito\discreteparticlenoise}_{\text{dry}},
    \label{eqn:dry}
\end{align}
where $D^{\rm dry}$ is a matrix of dry diffusion coefficients and $\discreteparticlenoise$ is a vector of independent Gaussian processes. The second dry diffusion term, $\nabla_x \cdot D^{\rm dry}$, is again an It\^{o} correction but in this case it is straightforward to compute directly. We define a ``wet percentage" as the percentage of the wet diffusion coefficient to the total diffusion coefficient of the particles $D^{\rm wet}/D^{\rm tot}$; as discussed above $D^{\rm wet}$ is determined from the effective hydrodynamic radius of the particles (depending on the grid spacing and Peskin kernel), and $D^{\rm tot}$ is prescribed as a particle property (note that $D^{\rm dry}=D^{\rm tot}-D^{\rm wet}$).

\subsection{Implementation of a mixed-type boundary}\label{subsec:bc}
\begin{figure}[h!]
    \centering
    \includegraphics[width=0.95\textwidth, trim={0cm 7.5cm 0cm 0cm}, clip]{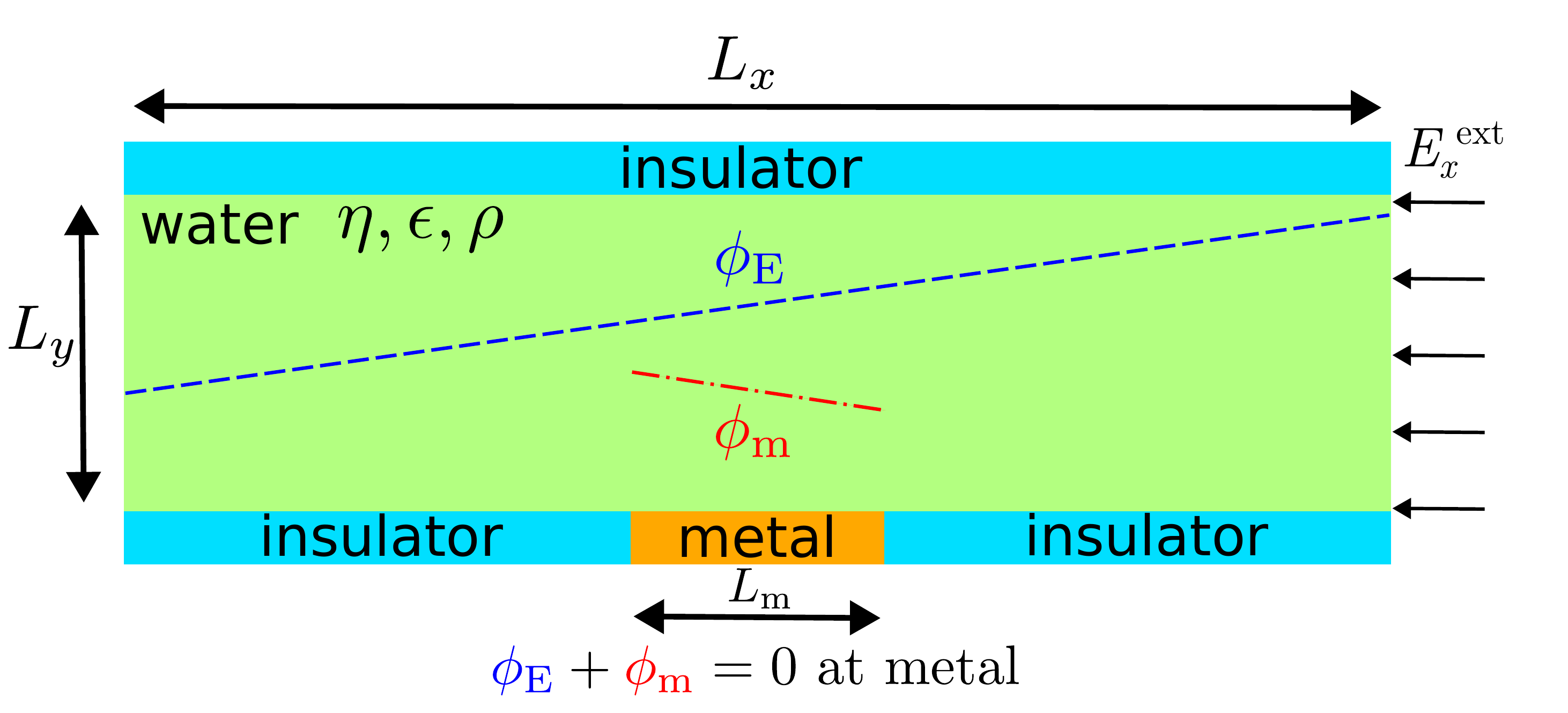}
    \caption{Cross-section of the three-dimensional simulation setup (not drawn to scale) and an illustration on the boundary conditions to solve electrostatic Poisson equation Eq.~(\ref{eqn:poisson}). The external electric field corresponds to a linear potential $\phi_E$. To ensure a homogeneous Dirichlet boundary for potential at the metal surface, we need to account for $\phi_E$ in specifying boundary conditions for the potential $\phi_m$. Specifically, we set $\phi_m = -\phi_E$ on the conducting part of the boundary.
    }\label{fig:setup}
\end{figure}

To simulate ICEO phenomena, we consider a doubly-periodic domain in $x$ and $z$-directions with walls in $y$-direction. A diagram illustrating the setup is present in Fig.~\ref{fig:setup}. For hydrodynamics, both top and bottom walls are no-slip, impenetrable, stationary walls. For electrostatics, the boundary conditions are more complex because we need to model a metal strip in the middle of the bottom wall, while the top wall is everywhere dielectric. Mathematically, the (ideal) metal conductor corresponds to a homogeneous Dirichlet boundary condition where electric potential vanishes at the surface (infinite permittivity). The (ideal) dielectric surface corresponds to a homogeneous Neumann boundary condition where surface charge is zero (zero permittivity).\footnote{This assumes that the permittivity of the fluid is much greater than that of the surface -- given the relatively large permittivity of water this is typically a good approximation. See Ref.~\citenum{maxian2021fast} for an approach for implementing non-zero boundary permittivity.}

One factor to consider regarding the electrostatic boundary condition is the uniform external applied electric field. The external field corresponds to an external electric potential, $\phi_E$, with a constant gradient parallel to the plate $d\phi_E/dx = -E_x^{\text{ext}}\hat{x}$. 
This external potential needs to be accounted for in solving for the full electrostatic potential.
The approach we have used is to solve Eq. (\ref{eqn:poisson}) for a correction potential, $\phi_m$ such that $\phi = \phi_E + \phi_m$.
We then set $\phi_m = -\phi_E$ on the metal plate to guarantee that the full potential $\phi$ satisfies a homogeneous Dirichlet boundary condition on the plate.

\subsection{Numerical parameters}\label{subsec:param}
We conducted our ICEO simulation in a channel of height $L_y=6.59$~nm in the $y$-direction, discretized on a 48-cell grid with the 4-point Peskin kernel to achieve 75\% wet percentage (see Appendix~\ref{appx:dry} for a discussion of this choice and its impact on the results). 
The parameters of the six cases we explored are summarized in Table \ref{table:param}. 
\begin{table}[t]
    \begin{ruledtabular}
    \begin{tabular}{ |c|c|c|c|c|c|c| }
        Cases & A & B & C & D & E & F \\ 
        \hline
        $E_x^{\textrm{ext}}$ (V/cm) & $5 \times 10^5$ & $1 \times 10^6$ & $1.5 \times 10^6$ & $2.5 \times 10^6$ & $5 \times 10^6$ & $1 \times 10^7$ \\ 
        $L_x$ (nm) & 26.36 & 26.36 & 52.72 & 52.72 & 105.44 & 210.88 \\ 
        $N_{cell,x}$ & 192 & 192 & 384 & 384 & 768 & 1536 \\
        $N_{+}(=N_{-})$, 0.15~M & 400 & 400 & 800 & 800 & 1600 & 3200 \\
        $N_{+}(=N_{-})$, 1~M & 2500 & 2500 & 5000 & 5000 & 10000 & 20000 \\
    \end{tabular}
    \caption{Parameters of our simulations}
    \label{table:param}
    \end{ruledtabular}
\end{table}
Note that for all the cases we study in this work, the average ions per cell (less than 0.00045 ions per cell) is so low that the continuum assumption can be problematic in methods like PNP and FHD. All cases use the same mesh spacing of $\Delta x=\Delta y=\Delta z=0.137$~nm in $x$-, $y$- and $z$-directions. The dimension in $z$-direction is fixed at $L_z=26.36$~nm and the number of cells in $z$-direction is $N_{cell,z}=192$. 
The channel length in the $x$-direction ($L_x$) depends on the electric field, reflecting the increased size of the vortices at high electric fields, which is briefly discussed in Sec.~\ref{subsec:velrhoe}. 
For all the cases explored, the metal strip has a fixed length $L_m=5.272$~nm and is centered on the lower boundary; see Fig.~\ref{fig:setup}. 
The time step used in all cases is 0.1~ps, which is much less restrictive than the typical molecular dynamics simulations that is on the order of femtosecond. The effective total simulation time for each case is on the order of 0.1--1~$\mu$s, which takes ~3,000 node-hours per case on NERSC Perlmutter supercomputer. 
The parameters describing the electrolyte composition are as follows: the system is a 1:1 electrolyte solution with $N_+$ cations and $N_-$ anions ($N_+=N_-$) at temperature $T=300$~K. 
The cation and anion charges are $q_{+}=-q_{-}=1.6 \times 10^{-19}$~C, and their diffusion coefficients are the same $D^\text{tot}_A = D^\text{tot}_C = 1.89\times 10^{-5}$~$\text{cm}^2\text{s}^{-1}$. The solvent is water with viscosity $\eta=9\times10^{-3}$~$\mathrm{g/(cm \cdot s)}$ and relative permittivity $\epsilon_r=80$. The ion-ion interactions use a 12-6 LJ potential (Eq.~\ref{eqn:LJ126}) with $\xi=8.16 \times 10^{-22}$~J and $\sigma=0.442$~nm. The ion-wall interactions use a 9-3 LJ potential (Eq.~\ref{eqn:LJ93}) where $\xi_{\textrm{wall}}=7.95 \times 10^{-21}$~J and $\sigma_{\textrm{wall}}=0.426$~nm. The bulk concentrations of the electrolyte $c^0=c_+^0=c_-^0$ we test are 0.15~M and 1~M. The relevant length scale in the electrolyte solutions is the Debye length $\Dlength$:
\begin{equation}
    \Dlength = \left(\frac{q_+ \chargedensity_+^0 + q_- \chargedensity_-^0}{\epsilon \kb T} \right)^{-1/2},
    \label{eqn:lambda_D}
\end{equation}
where $\chargedensity_{\pm}^0$ are the charge densities of the species in the bulk. Eq.~\ref{eqn:lambda_D} gives the Debye length of 0.796~nm and 0.319~nm for the cases of 0.15~M and 1~M, respectively.

\section{Results and discussions}\label{sec:results}
We extend our previous study \cite{ladiges2022ekdiscos} of electro-osmosis using DISCOS to induced-charge electro-osmosis, a more complex configuration that involves mixed-type boundaries. Our main goal is to explore the velocity scaling in ICEO with respect to a wide range of electric fields at moderate electrolyte concentrations, and compare with other simulation methods and theory. 
The results suggest a new power-law velocity scaling at moderate to high electric fields. We propose possible explanations for the discrepancies with theory.

\begin{figure}[h!]
    \centering
    \includegraphics[width=\textwidth, trim={0 6.5cm 0 0}, clip]{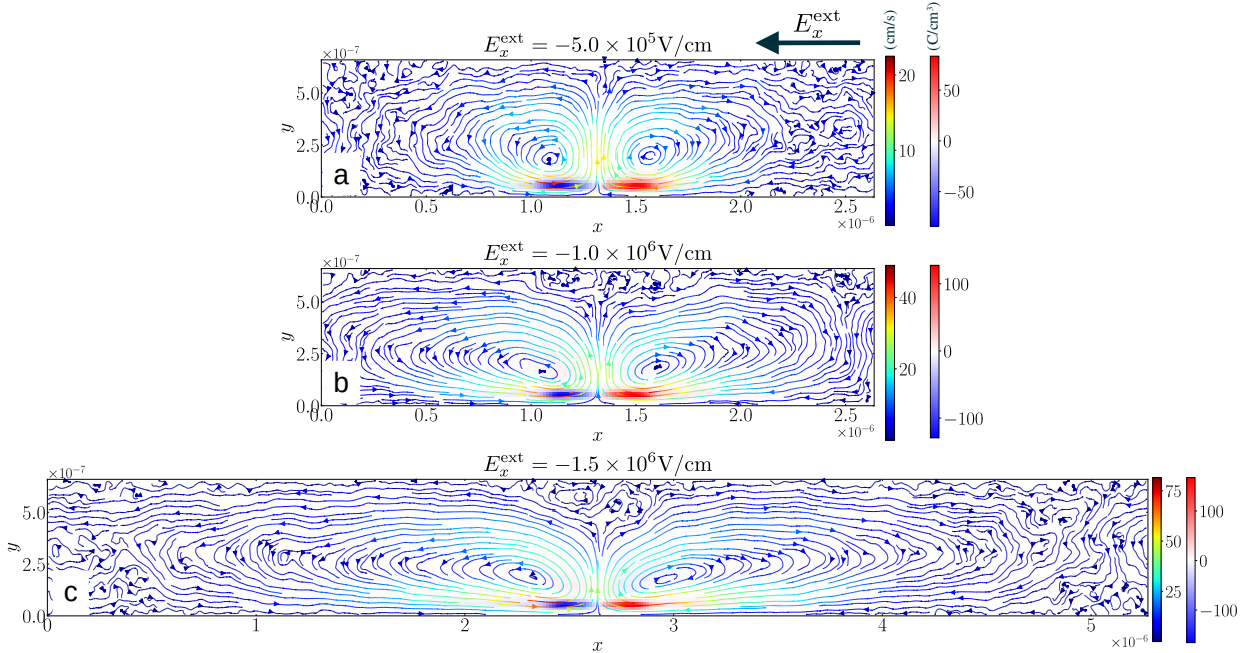}
    \caption{Time-averaged flow properties and charge density distributions for DISCOS simulations of (a) Case~A, (b) Case~B, and (c) Case~C for 0.15~M.
    Left and right color bars correspond to flow speed and charge density, respectively.
    Note that Case~C uses a larger domain in $x$-direction because it helps eliminate the finite system size effect. For higher electric fields we use even larger domains in the $x$-direction; see Table~\ref{table:param}. }
    \label{fig:flow}
\end{figure}
\subsection{Flow field and charge distribution}\label{subsec:velrhoe}
\begin{figure}[h!]
    \centering
    \includegraphics[width=\textwidth, trim={0 5.5cm 2cm 0}, clip]{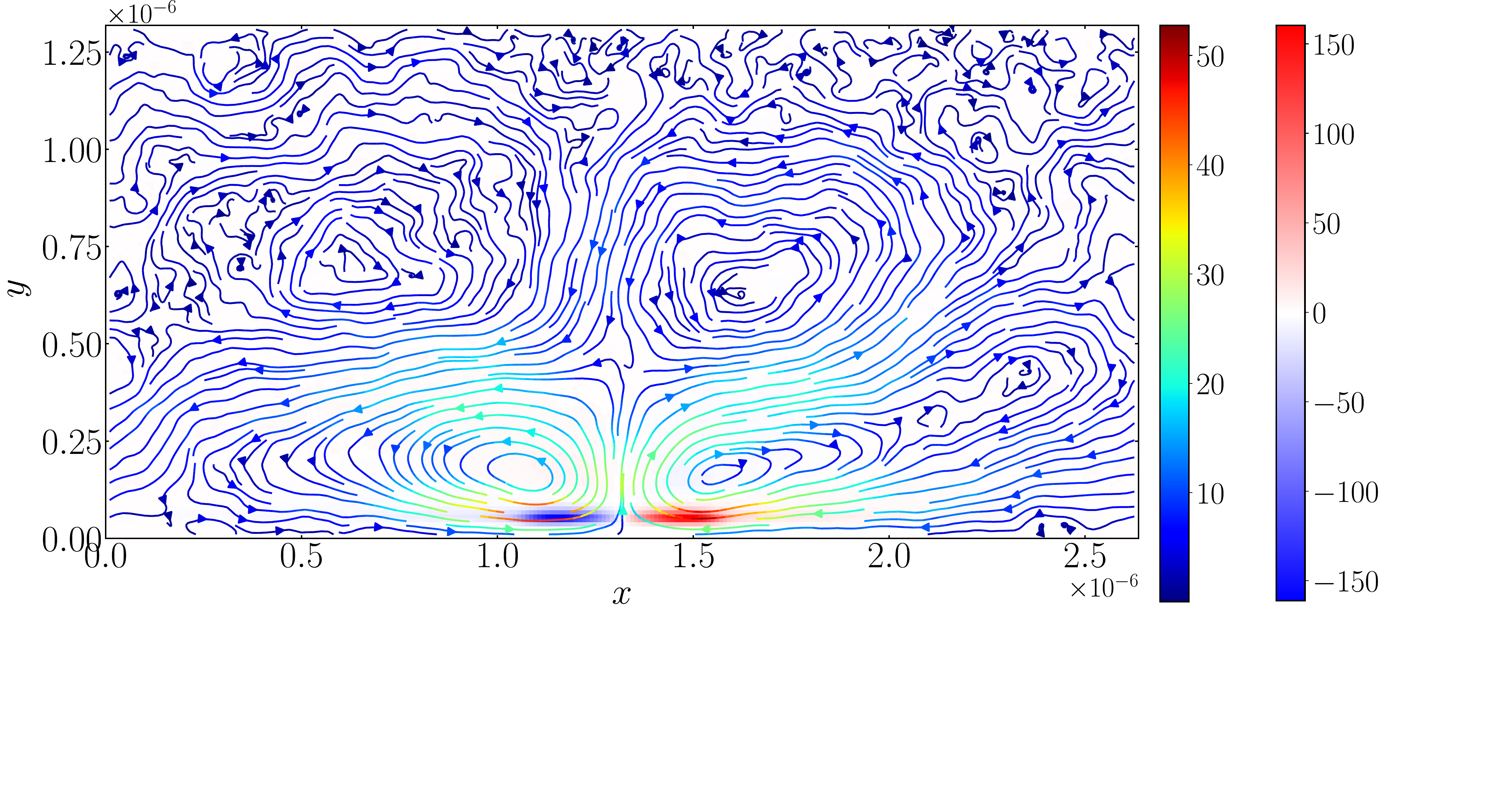}
    \caption{Time-averaged flow properties and charge density distributions for DISCOS simulations of Case~B for 0.15~M, but with a 13.18~nm-high channel. See Fig.~\ref{fig:flow}(b) for comparison.}
    \label{fig:flow_13nm}
\end{figure}

We first validate that our simulations qualitatively capture the ICEO behavior by presenting time-averaged flow field and charge distribution results.\footnote{For an instantaneous plot, which appears to be quite noisy, see Ref.~\cite{ladiges2022ekdiscos}.} Figure~\ref{fig:flow} shows flow streamlines superimposed with a raster plot of charge density for Cases~A, B, and C at 0.15~M obtained from the DISCOS algorithm; the left and right color bars correspond to flow speed and charge density, respectively. The flow pattern and charge distribution match our expectations: a counter-rotating vortex pair is formed, with cations accumulating on the right side of the metal plate and anions accumulating on the left, due to a right-to-left electric field. 
We repeated the simulations Cases~A, B, and F, which include the lowest and highest fields we studied, in a channel with twice the height ($L_y=13.18$~nm). Case B in the 13.18~nm-high channel is shown in Fig.~\ref{fig:flow_13nm} and the primary flow pattern near the bottom does not change noticeably (in spite of the weak vortices near the top wall due to the no-slip and no-penetration boundary conditions for the system), so the upper wall for the 6.59~nm-high channel has a minimal effect on the flow field. 
We observe that the ions accumulate roughly one $\sigma$ (Lennard-Jones size) above the bottom wall, illustrating the steric effect captured by DISCOS. The location of the peak velocity also occurs roughly near the location where ions accumulate, which is expected since the breaking of electroneutrality in the double layer drives the flow.

\begin{figure}[t!]
    \centering
    \includegraphics[width=0.95\textwidth, trim={0 8.5cm 0 0}, clip]{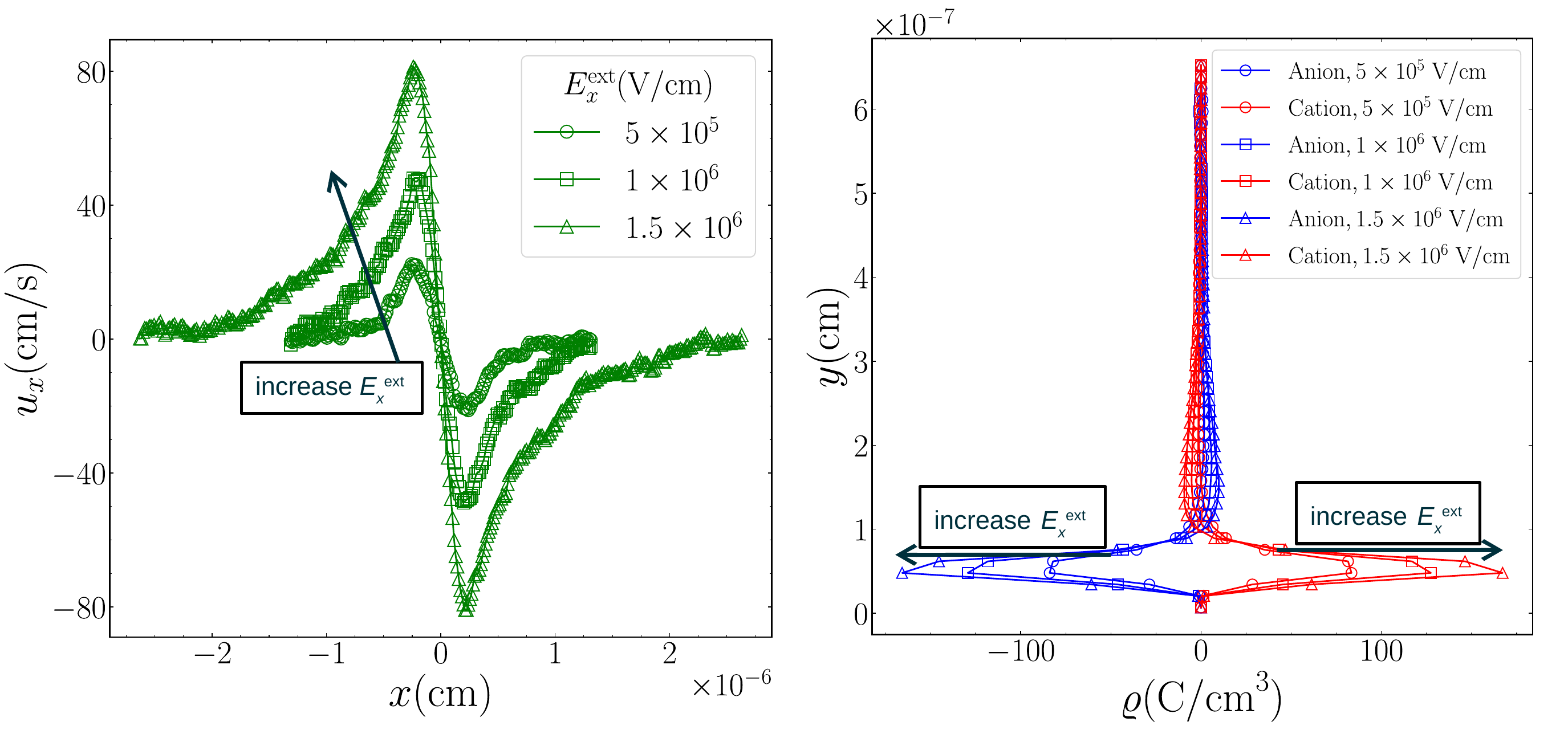}
    \caption{Horizontal fluid velocity and ion densities for Cases A, B and C at 0.15~M.
    Left: time-averaged $x$-velocity as a function of $x$ at $y=0.618$~nm where peak velocity occurs. The $x$-coordinates are shifted by $L_x/2$ of the corresponding case so that the centers of the metal plate are aligned.
    Right: time-averaged charge density as a function of $y$ at $x$ locations where maximum charge accumulation occurs (at $x=11.33$ and $15.03$~nm for Case~A, 
    at $x=11.46$ and $14.90$~nm for Case~B, 
    and at $x=24.78$ and $27.94$~nm for Case~C). 
    }
    \label{fig:peak_vel_rhoE}
\end{figure}
\begin{figure}[t!]
    \centering
    \includegraphics[width=0.95\textwidth, trim={0 8cm 0cm 0cm}, clip]{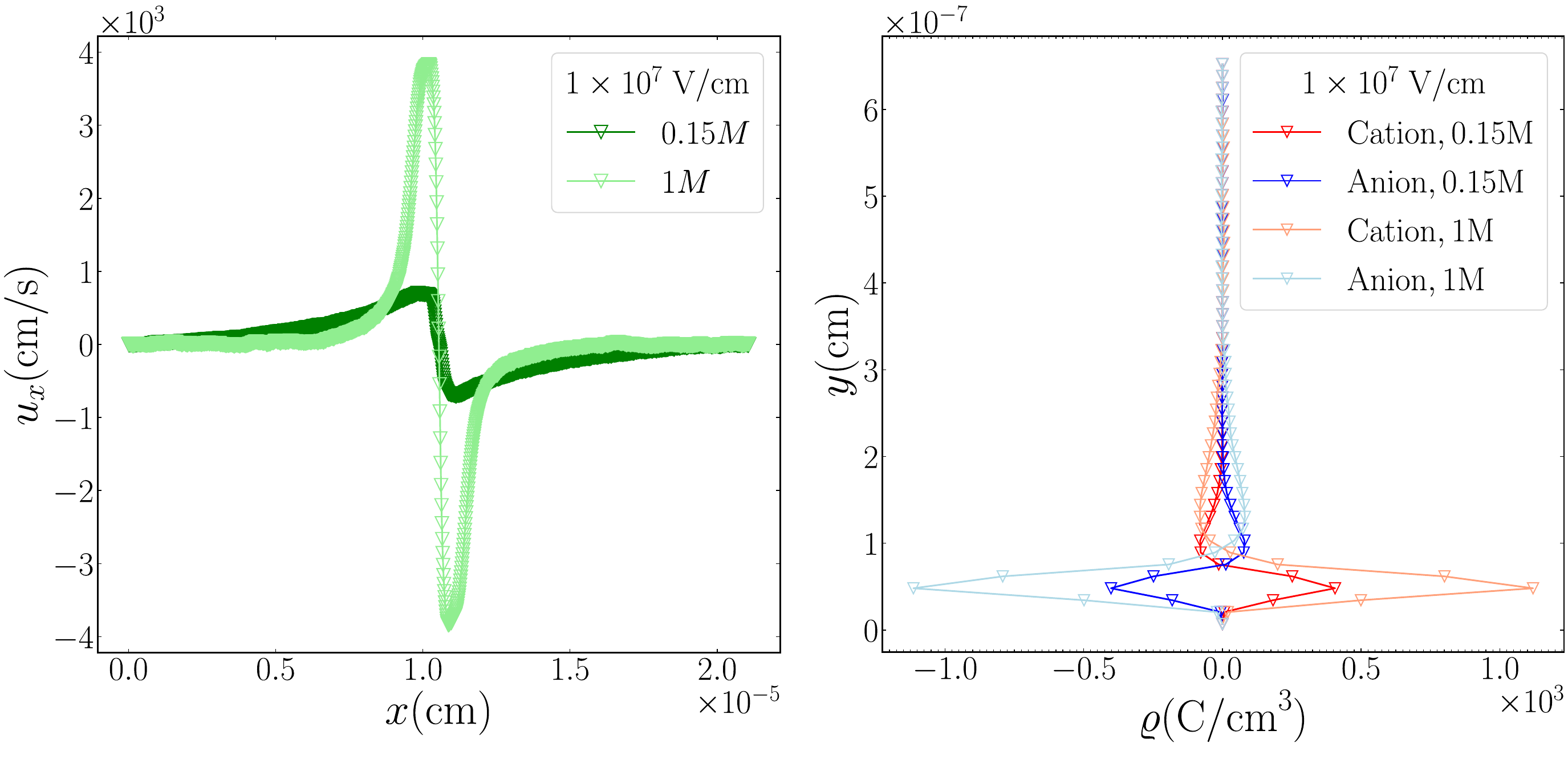}
    \caption{Flow properties and charge distribution for Case F ($E_x^{\textrm{ext}}=1 \times 10^7$~V/cm) at 0.15~M and 1~M. Left panel: time-averaged $x$-velocity profile in $x$ at $y=0.618$~nm where peak velocity is obtained. Right panel: time-averaged charge density profiles in $y$ at $x=104.72$ and $106.16$~nm where maximum charge density is reached at this field.}
    \label{fig:1e14}
\end{figure}
The flow and charge patterns show several intriguing features that depend on the electric field strength (Fig.~\ref{fig:flow} from top to bottom panels):

(1) At higher fields ($E_x^{\textrm{ext}} \geq 1.5 \times 10^6$~V/cm), we observe wider vortices for which we increase the domain length in the $x$-direction accordingly as noted in Table~\ref{table:param} to avoid finite system size effects from periodic boundaries. This elongation of the vortices is due to a stronger fluid velocity, as revealed in the increasing peak velocity with electric field in the left plot of Fig.~\ref{fig:peak_vel_rhoE}, which presents the $x$-velocity profile along the horizontal slices where peak velocity occurs.

(2) Under the external field there is a discontinuity of (induced) surface charge density at the edge of the metal where the development of double layer occurs, thus driving the ICEO flows. 
The $x$-location of the peak velocity above the plate occurs near the edge of the metal plate in all of the simulations. 
We remark that the peak velocities in our simulation ($\sim$cm/s) are significantly higher than those reported in experiments ($\sim \mu$m/s) \cite{brown2000pumping,green2000fluid,green2002fluid,studer2004acekpump,levitan2005experimental,ramos2005pumping,bown2006aceo,urbanski2006fast,urbanski2007effect,bazant2007electrolyte,soni2007iceo,storey2008steric}.
One potential explanation of this discrepancy is that we simulate a direct-current (DC) field that continuously charges the double layer above the metal plate with ions, while most experiments use an alternating-current (AC) field that involves charging and discharging, resulting in a weaker flow field. 
The Helmholtz-Smoluchowski equation \cite{helmholtz1879mobility,lyklema2003ek} predicts that a stronger electric field enhances the charge accumulation, as shown in the charge density plot of Fig.~\ref{fig:peak_vel_rhoE} (right panel), which leads to a faster flow. 

(3) We observe an overcharging effect at higher electric field strengths ($E_x^{\textrm{ext}} \geq 1.5 \times 10^6$~V/cm), that is, an additional layer of co-ions to the surface charge starts to emerge above the primary counterion layer. This effect seemingly becomes more pronounced with increasing electric field due to the steric effect, demonstrating the importance of these molecular-scale effects in nanoscale electrokinetic phenomena.
Figure~\ref{fig:1e14} shows that these trends of increasing velocity, growing charge density and the overcharging effect persist for the highest electric field we tested (Case F, $E_x^{\textrm{ext}}=1 \times 10^7$~V/cm at 0.15~M (dark-colored curves)). The simulation results for 1~M show qualitatively similar but higher flow velocities and charge density, as expected; see Fig~\ref{fig:1e14} (light-colored curves). 

\begin{figure}[t!]
    \centering
    \includegraphics[width=0.95\textwidth, trim={0 12.8cm 0cm 0cm}, clip]{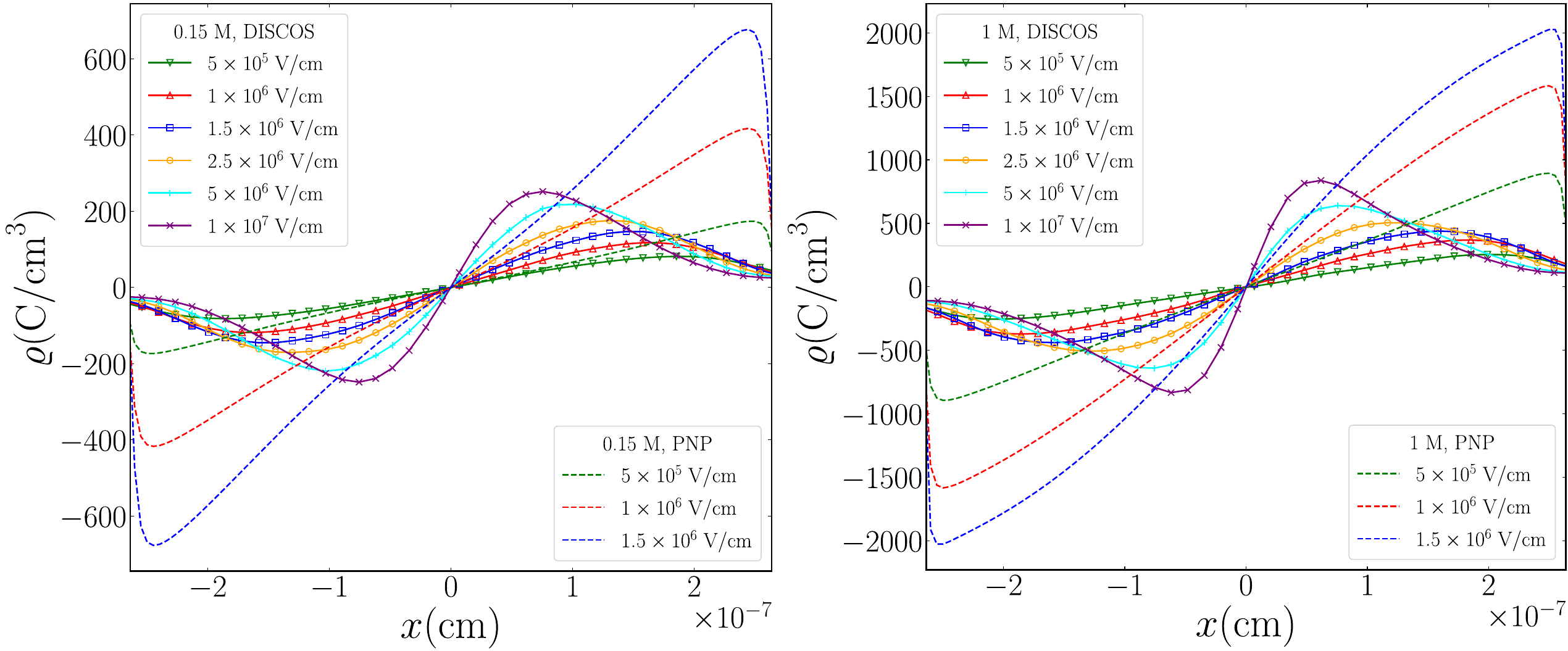}
    \caption{Net charge density profile at 0.15~M (left) and 1~M (right). The DISCOS results are time averaged. The data is measured at the peak charge density height: at $y=0.618$~nm (5 cells, which coincides with the height of the peak horizontal velocity) for DISCOS and at $y=0.0686$~nm (1 cell) for PNP. The $x$-coordinates are shifted by $L_x/2$ of the corresponding case so that the centers of the metal plate are aligned. The abscissas exactly span the metal plate in the $x$-direction, so the left and right ends of both plots correspond to left and right edges of the metal plate.}
    \label{fig:rhoE_slice}
\end{figure}

(4) Figure \ref{fig:rhoE_slice} shows the charge density profile at the height where the peak charge density is achieved in both DISCOS and PNP simulations. First, note that in DISCOS the location of the maximum charge density is midway between the edge and the center of the plate while it is near the edges of the plate in PNP. We further find that in DISCOS this peak shifts slightly toward the center of the metal plate with increasing electric fields.
Consequently DISCOS simulations exhibit disparity in the $x$-location of the maximum charge density and the $x$-location of the maximum horizontal velocity (which always occur near the edge of the metal plate, Fig.~\ref{fig:peak_vel_rhoE} left plot). 
We also observe that DISCOS predicts the maximum charge occurs at a height of 0.618~nm 
above the wall whereas for PNP it occurs at 0.0686~nm, in the first computational cell. 
Steric effects are again a contributor to these phenomena, which is not present in the PNP system where ion concentration can become arbitrarily large in the vicinity of the edge of the metal plate (see Fig.~\ref{fig:rhoE_slice}). 

\subsection{Velocity scaling}\label{subsec:scaling}
One of the most fundamental criteria to determine the efficiency of electrokinetic flows is the velocity scaling with respect to the electric field, describing how effectively electrical energy is converted into mechanical flow. The higher the scaling power, the more effective the conversion is. Theoretically in the low field strength regime for regular electro-osmotic (EO) flows, the slip velocity is described by the Helmholtz-Smoluchowski formula \cite{helmholtz1879mobility,lyklema2003ek}:
\begin{equation}
    v_s = \frac{\epsilon \phi_{\zeta}}{\eta} E_{\parallel},
    \label{eqn:iceo_vel}
\end{equation}
where $\phi_{\zeta}$ is the potential change across the double layer, commonly referred to as the $\zeta$-potential, and $E_{\parallel}=E_x^{\textrm{ext}}$ is the electric field parallel to the boundary. In ICEO $E_\parallel$ is given by the external field $E_x^{\textrm{ext}}$ and the $\zeta$-potential in the low field strength regime is proportional to the external field ($\phi_\zeta \sim E_{\parallel} L_m$), that is,
\begin{equation}
    v_s = \frac{\epsilon L_m}{\eta} E_{\parallel}^2,
    \label{eqn:iceo_vel2}
\end{equation}
so the overall characteristic velocity in ICEO scales quadratically with external field $v_s \sim E_{\parallel}^2$, which is more favorable than the linear scaling in regular EO. 

\begin{figure}[b!]
    \centering
    \includegraphics[width=0.5\textwidth, trim={0cm 0 1cm 1cm}, clip]{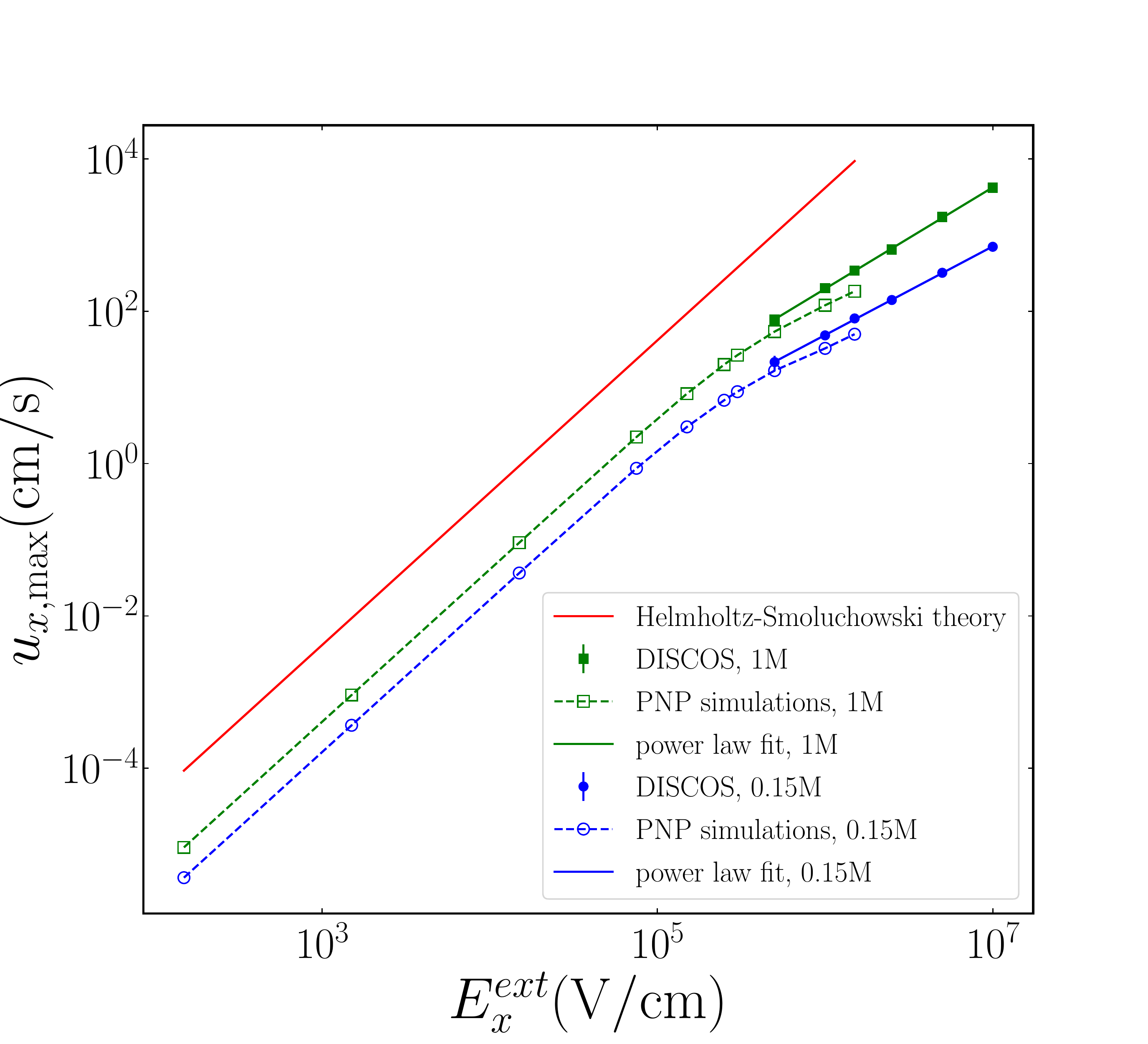}
    \caption{Peak velocity measured from DISCOS (time-averaged) and PNP as a function of external electric fields, compared with theory (Eq.~(\ref{eqn:iceo_vel2})). The error bars represent one standard error, which is so small that they are smaller than the size of the symbols.}
    \label{fig:scaling}
\end{figure} 
In our ICEO simulations, we explore how the fluid velocity scales with electric field, and compare those measurements to the quadratic scaling, using the time-averaged peak velocity as the characteristic velocity. Figure~\ref{fig:scaling} shows the time-averaged peak $x$-velocity as a function of electric field from our DISCOS simulations for 0.15~M and 1~M, as well as the Poisson-Nernst-Planck (PNP) based, continuum hydrodynamics simulations and the theoretical prediction, Eq.~(\ref{eqn:iceo_vel2}). A detailed comparison between the DISCOS simulation and the PNP model for 1.0~MV/cm (Case B) was presented in our previous paper \cite{ladiges2022ekdiscos}.  Here we extend both DISCOS and PNP simulations to a wide range of electric fields, but in different regimes. Fig~\ref{fig:scaling} shows PNP results for weak ($< 0.5$~MV/cm) to intermediate ($0.5 - 1.5$~MV/cm) electric fields, and DISCOS results for intermediate to strong ($> 1.5$~MV/cm) electric fields. The different ranges reflect the differences in applicability of the two methods. 
The PNP equations coupled with the Navier-Stokes equations is effective for relatively weak fields \cite{newman2012electrochemical}. This approach has also been extended to incorporate thermal fluctuations \cite{peraud2016low,peraud2017fluctuation}. However, the PNP equations begin to break down as the electric field strength is increased, leading to high concentrations at boundaries.  This problem is particularly noticeable in ICEO because of a singularity in the electric field arising from the transition from Neumann to Dirichlet boundary conditions for the electrostatic equation at the edge of the metal plate.
The singularity scales as $r^{-1/2}$ as the transition point is approached, where $r$ is the distance from the transition. As the field strength is increased, the PNP equations predict an increase in ion concentration at the boundary where it interacts with the singularity in a nonphysical fashion.
On the other hand, DISCOS incorporates steric effects that provide a more realistic model for ions.  As a result, ions are kept away from the singularity point, thus allowing DISCOS to simulate much higher electric fields; see discussions in Sec.~\ref{subsec:steric}.
However, for weak electric fields where mean fluid velocity signal is comparable to that of the thermal fluctuations, DISCOS needs to average over an extremely large ensemble to get good statistics, making the computational cost prohibitively expensive. 

We also remark that our high-molarity case (1~M) shows higher velocities than the low-molarity case (0.15~M), in contrast to experiments \cite{gangwal2008induced,bazant2009towards}.
One possible explanation of this discrepancy is that our simulations employ DC fields that induce a persistent double layer, while the experiments used AC fields that involve a charging-discharging process. 
In the DISCOS simulations, the charge density inside the double layer is nearly constant in time and 
should be proportional to the bulk electrolyte concentration, while the Debye length scales inversely with the square root of concentration (Eq.~\ref{eqn:lambda_D}). 
In fact, the charge density ratio observed in Fig.~\ref{fig:rhoE_slice} between 1~M and 0.15~M is only around 3, smaller than the bulk concentration ratio of $1~\mathrm{M}/0.15~\mathrm{M}=6.67$, but it still compares favorably with the estimated ratio of Debye lengths $\sqrt{1.0~\mathrm{M}/0.15~\mathrm{M}}=2.58$.
Thus the total charge within the double layer increases with increasing molarity, resulting in a growth of velocity. 

Intriguingly, we observe an almost linear velocity scaling with electric field in DISCOS simulations when $\zeta$-potential is beyond the thermal voltage. The blue solid line in Fig.~\ref{fig:scaling} is a power-law fit to the mean values of DISCOS results (blue dots), giving a scaling slightly larger than 1, $u_{x,\textrm{max}} \sim |E_x^{\textrm{ext}}|^{1.16}$. The result at 1~M shows a similar trend as 0.15~M, in which the deviation from $(E_x^{\textrm{ext}})^2$ scaling also emerges, with a power law fit of $u_{x,\textrm{max}} \sim |E_x^{\textrm{ext}}|^{1.33}$.
Such phenomenon of suppressing power to the velocity scaling has been derived in a theory regarding dielectric spheres/cylinders with zero surface charge \cite{schnitzer2014strong} and has been observed both in ICEO experiments with metal/Janus spheres \cite{peng2014induced,feng2020recent} and in simulations with polarizable cylinders \cite{davidson2014chaotic}, but it has not been discussed systematically in planar ICEO flows. 
The PNP simulations show a transition in the velocity scaling as $\zeta$-potential exceeds the thermal voltage. It faithfully captures quadratic velocity scaling at small fields (Eq.~\ref{eqn:iceo_vel}), although quantitatively the results are an order of magnitude smaller than the Helmholtz-Smoluchowski theory.
This quantitatively lower ICEO velocity is not surprising as it is well-known that the theory typically over-estimates the ICEO velocity by one or two orders of magnitude compared to experiments with various geometries at much lower concentrations than our simulations
\cite{brown2000pumping,green2000fluid,green2002fluid,studer2004acekpump,levitan2005experimental,ramos2005pumping,bown2006aceo,urbanski2006fast,urbanski2007effect,bazant2007electrolyte,soni2007iceo,storey2008steric,bazant2009towards,feng2020recent}. 
The velocity scaling then gradually reduces to a smaller scaling power, seemingly matching the scaling law of DISCOS results. 
For electric field strengths (Case A, B and C) where both DISCOS and PNP simulations can be conducted, the peak velocities from the two methods differ slightly, with velocities obtained from DISCOS always larger than those obtained from the PNP method. 
The difference between the peak velocities from the two methods widens as electric field increases, again demonstrating the importance of the molecular-scale phenomena such as the steric effect in this regime. 
In the next section, we discuss the possible mechanisms that can play a role in this velocity scaling turnover, and remark on their implications for broader ICEO applications.

\subsection{Effect of charge-induced thickening}\label{subsec:thickening}
Many hypotheses have been proposed to explain the suppressed velocity scaling in ICEO compared to theory, which are nicely summarized in a review by Bazant {\it et al.} \cite{bazant2009towards}. 
Here we focus on one specific hypothesis called charge-induced thickening \cite{bazant2009towards}, which can be systematically tested in our simulations. This effect is postulated from the fact that the electrical double layer is highly concentrated with the counterions to the surface charge at high electric fields, so the crowding of the counterions leads to an increase in the local viscosity. The theory predicts a power-law dependence of the ratio of electric permittivity and viscosity as the charge density approaches a critical value:
\begin{equation}
    \frac{\epsilon}{\eta} = \frac{\epsilon_b}{\eta_b}\left[1-\left(\frac{|\rho|}{\rho^{\pm}_j}\right)^{\alpha}\right]^{\beta},
    \label{eqn:eo_ratio}
\end{equation}
where $\epsilon_b$ and $\eta_b$ are electric permittivity and viscosity of the bulk, $|\rho|$ is the charge density, $\rho^{\pm}_j$ is the critical (positive or negative) charge density of species $j$, and $\alpha$ and $\beta$ are critical exponents. Eq.~\ref{eqn:eo_ratio} is equivalent to a power-law divergence of viscosity assuming permittivity is constant throughout the system $\epsilon = \epsilon_b$:
\begin{equation}
    \eta  = \eta_b \left[1-\left(\frac{|\rho|}{\rho^{\pm}_j}\right)^{\alpha}\right]^{-\beta},
    \label{eqn:visc_fit}
\end{equation}
The ratio in Eq.~\ref{eqn:eo_ratio} is linked to a general electro-osmotic mobility $b$ \cite{lyklema1995fundamentals}:
\begin{equation}
    b=\int^{\Psi_D}_0 \frac{\epsilon}{\eta} d\Psi,
    \label{eqn:eo_mob}
\end{equation}
where $\Psi_D$ is the potential drop across the diffuse layer (i.e. the potential difference between the surface and the bulk). 
The mobility $b$ defines an effective $\zeta$-potential:
\begin{align}
    \phi_{\zeta,\textrm{eff}} &= b\frac{\eta_b}{\epsilon_b}\label{eqn:eff_zeta1} \\ 
    &= \int^{\Psi_D}_0 \left[1-\left(\frac{|\rho|}{\rho^{\pm}_j}\right)^{\alpha}\right]^{\beta} d\Psi\label{eqn:eff_zeta2} \\ 
    &= \int^{L}_0 \left[1-\left(\frac{|\rho(y)|}{\rho^{\pm}_j}\right)^{\alpha}\right]^{\beta} \frac{d\Psi}{dy} dy, \label{eqn:eff_zeta3}
\end{align}
where in Eq.~\ref{eqn:eff_zeta1} and Eq.~\ref{eqn:eff_zeta2} $\epsilon = \epsilon_b$ is assumed. The equations above allow us to construct a modified Helmholtz-Smoluchowski formula in the modified Poisson-Boltzmann model \cite{bazant2009towards}. Physically, the effect of Eq.~\ref{eqn:eo_ratio} is to generate a thick diffuse layer of ions that extends beyond the electric double layer with slowly decaying ion concentration, at high electric fields. As a result, the $\zeta$-potential at high electric fields becomes smaller than $\Psi_D$, while at low fields the $\zeta$-potential and $\Psi_D$ are roughly the same. So overall, there are two key parameters governing the hypothesized charge-induced thickening effect: the viscosity and the $\zeta$-potential. We measure these two quantities next.

\begin{figure}[h!]
  \centering
    \includegraphics[width=0.5\textwidth, trim={0 0cm 0 0}, clip]{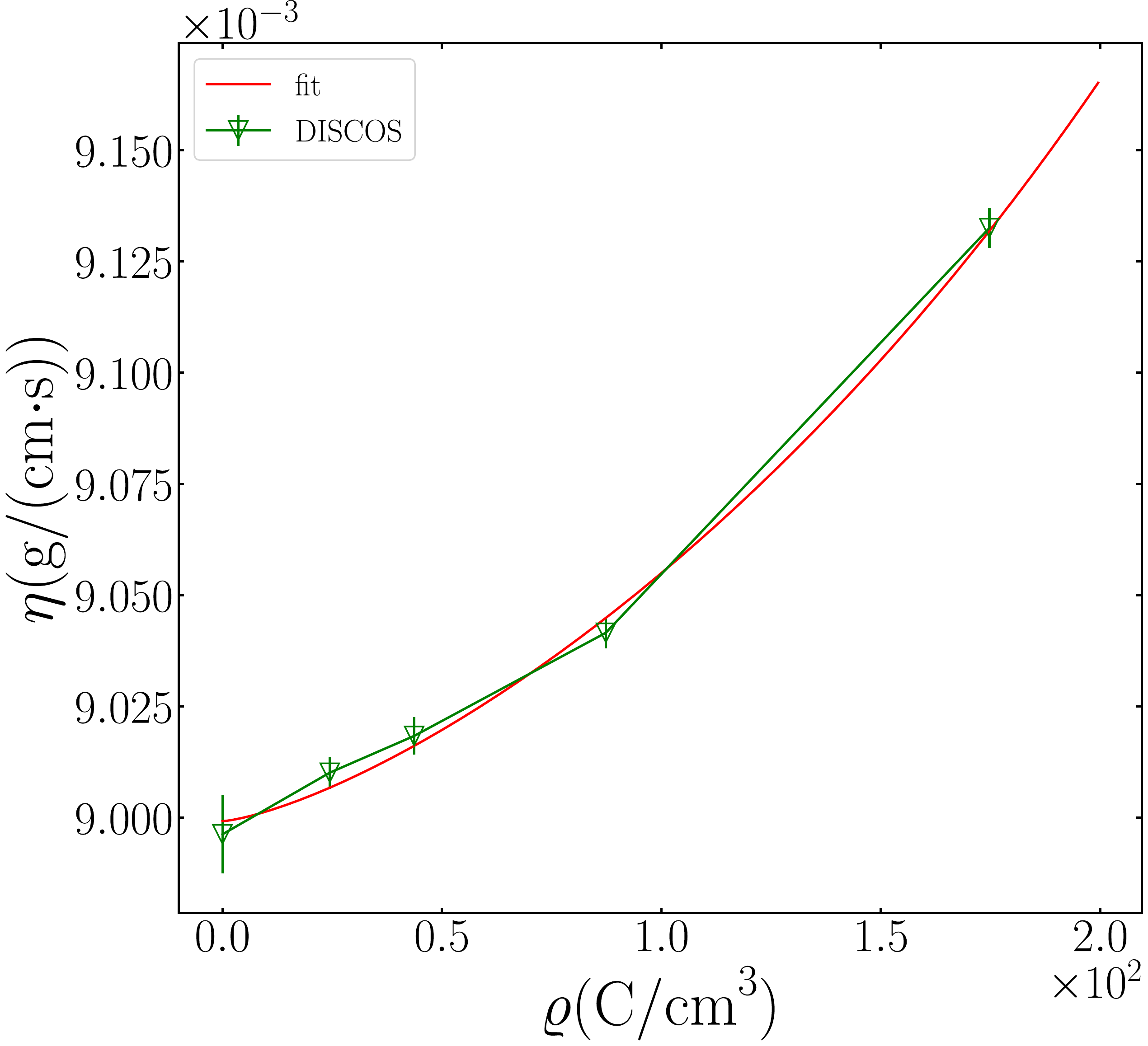}
    \caption{Measured viscosity from Poiseiulle flow in a channel of like-charged ions with different concentrations, using DISCOS method. A power-law fit of Eq.~\ref{eqn:visc_fit} is applied to the data points. }
    \label{fig:visc_discos}
\end{figure}
We first directly measure the system viscosity and examine whether it is the main contributor to the deviation from an $(E_x^{\textrm{ext}})^2$ velocity scaling. The viscosity is measured by a series of Poiseuille-flow DISCOS simulations of like-charged ions in a channel that has the same dimensions as Case A. It is again periodic in $x$- and $z$-directions, and the channel walls at $y$-boundaries satisfy no-slip and no-penetration conditions (Dirichlet boundary condition) for the fluid solver, and a constant surface charge condition (Neumann boundary condition) for the electrostatic solver. 
We apply a body force to the system (solvent and solutes) that is equivalent to a pressure gradient of $dP/dx = 1 \times 10^{16}$~Pa/m along the channel, and measure the peak velocity $v_m$. The viscosity of the system $\eta$ is then computed using
\begin{equation}
    \eta = \frac{(dP/dx) L^2}{8 v_m}. 
    \label{eqn:visc_poiseuille}
\end{equation}
DISCOS captures the growth of viscosity in an electrolyte solution due to the presence of ions, as shown in Figure~\ref{fig:visc_discos} that plots measured viscosity versus charge density in the channel at 0~M, 0.2625~M, 0.46875~M, 0.9375~M and 1.875~M. We fit our viscosity data to Eq.~\ref{eqn:visc_fit} (assuming $\epsilon = \epsilon_b$), with $\rho^{\pm}_j$ preset to $\rho^{\pm}_j=q/\sigma^3$ using the ion-ion Lennard-Jones length $\sigma=0.442$~nm to represent a characteristic length scale between ions at jamming, to obtain $\rho^{\pm}_j=1.853 \times 10^3$~$\textrm{C/cm}^3$, $\alpha=1.529$ and $\beta=0.531$. The small value of $\beta$ indicates
a weakly increasing viscosity, so that the viscosity only becomes significant higher at extremely high charge density. The charge density near the edge of the metal plate, which is the place where peak velocity is induced, is at most around 50~$\textrm{C/cm}^3$ for 0.15~M, and is at most around 180~$\textrm{C/cm}^3$ for 1~M (see Fig.~\ref{fig:rhoE_slice}). The viscosity at these charge densities is less than 1.5\% above the solvent viscosity, so the velocity is suppressed by at most 1.5\%, and it generally does not change the velocity scaling with respect to the electric field. We conclude that the charge-induced viscosity growth plays a minor role in the velocity scaling turnover in our simulation. 

\begin{figure}[h!]
    \centering
    \includegraphics[width=0.95\textwidth, trim={0 8cm 0 0}, clip]{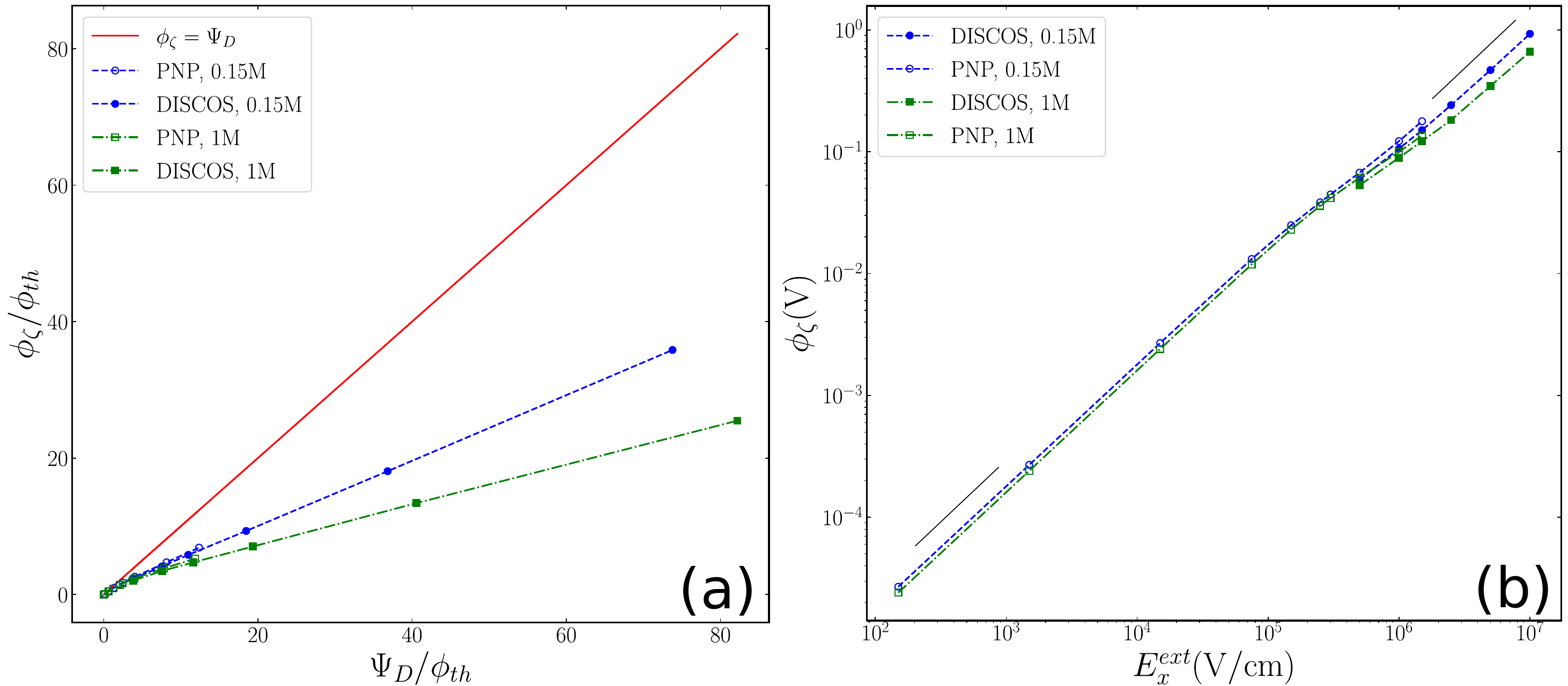}
    \caption{(a) Measured $\zeta$-potential $\phi_{\zeta}$ vs diffuse layer potential drop $\Psi_D$, compared with the ideal condition ($\phi_{\zeta}=\Psi_D$, solid red line). (b) Log-log plot of measured $\phi_{\zeta}$ as a function of applied electric field, which shows that $\phi_{\zeta}$ transitions from one linear regime to another linear regime with increasing electric field, indicated by the solid black lines. The location of the transition roughly coincides with the turnover of the velocity scaling.}
    \label{fig:zeta_psi}
\end{figure}
We next investigate the field-dependent $\zeta$-potential $\phi_{\zeta}$, which is the remaining term from Eq.~\ref{eqn:iceo_vel} that can affect the velocity scaling. Figure~\ref{fig:zeta_psi}(a) shows $\zeta$-potential versus $\Psi_D$ for different electric fields at two concentrations. Both horizontal and vertical axes are normalized by the thermal voltage $\phi_{th}=k_B T/e$ where $e$ is the elementary charge. We observe that (1) $\zeta$-potential is suppressed compared with the theory for ideal dilute conditions ($\phi_{\zeta} = \Psi_D$, red solid line) and (2) the suppression becomes more pronounced with increasing concentrations, both of which are reported in Ref.~\citenum{bazant2009towards} in the context of charge-induced thickening (Eq.~\ref{eqn:eff_zeta1}-\ref{eqn:eff_zeta3}). However, the physical origin of the suppression in our simulation is different from Ref.~\citenum{bazant2009towards}. In our simulations the suppression occurs because of a gradual potential drop across the channel due to the accumulation of ions at high fields forming an extended diffuse layer beyond the Debye length, without much influence from viscosity. As was discussed in the previous paragraph, charge-induced viscosity grows weakly, thus having a minor effect on the effective $\zeta$-potential.
Moreover, the measured $\zeta$-potential still scales linearly with $E_x^{\textrm{ext}}$, as shown in Fig.~\ref{fig:zeta_psi}(b), while theories \cite{schnitzer2012induced} have attributed the non-quadratic velocity scaling to a non-linear electric-field dependence of the $\zeta$-potential. The $\zeta$-potential in fact transitions from one linear regime in $E_x^{\textrm{ext}}$ to another linear regime in $E_x^{\textrm{ext}}$, with different pre-factors before and after the velocity scaling turnover. The linear scaling of $\zeta$-potential in $E_x^{\textrm{ext}}$ should still produce an $(E_x^{\textrm{ext}})^2$ scaling based on Helmholtz-Smoluchowski formula, yet we observe a significantly lower velocity scaling. This gap between the simulation and the prediction from Helmholtz-Smoluchowski formula suggests some fundamental factors are missing in the current theories at high electric fields, such as the steric effect. 

\subsection{Steric effect of ions}\label{subsec:steric}
\begin{figure}
    \centering
    \includegraphics[width=0.9\textwidth, trim={0 5cm 0 1.3cm}, clip]{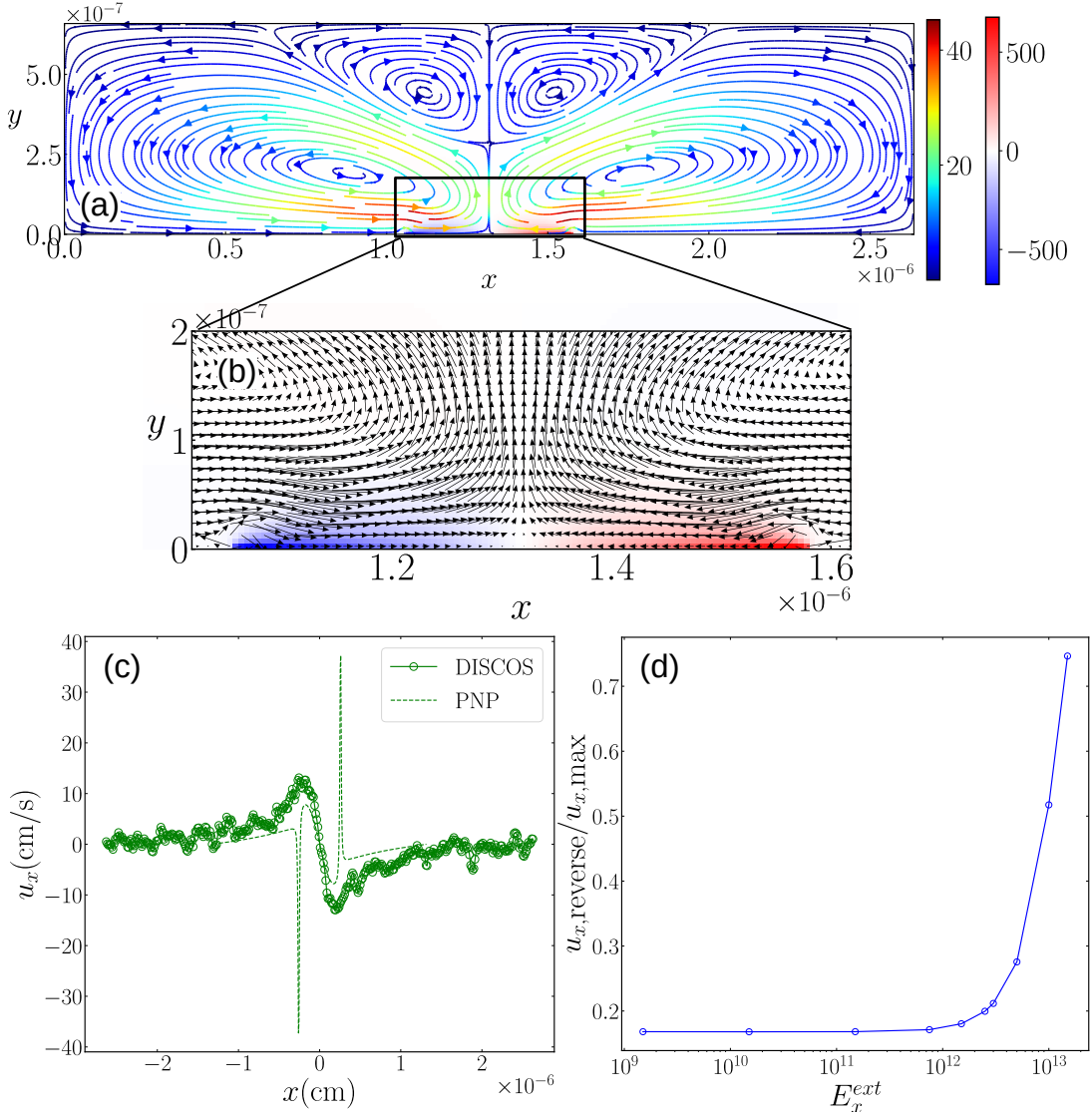}
    \caption{(a) Flow properties and charge density distribution for PNP simulations of Case~C at 0.15~M. The color bars represent fluid speed (left) and charge density (right), respectively. (b) Magnified view for the velocity vector field in the rectangular region indicated in (a). Notice that there are small vortices forming at the bottom boundary. (c) Horizontal velocity along $x$-direction just above the bottom boundary (at $y=0.0686$~nm, one cell from the boundary) for Case~C at 0.15~M, from DISCOS (time-averaged) and PNP methods. The $x$-coordinates are shifted by $L_x/2$ of Case~C. (d) The ratio of the reversed velocity and the peak velocity in the PNP simulation as a function of electric field. The reversed velocity is defined as the maximum velocity in the opposite direction; for example, the most negative velocity in the left half of the domain in (a) and the most positive velocity in the right half.
    }
    \label{fig:ux_reverse}
\end{figure}
One novel contribution that the DISCOS method provides is that it explicitly models the steric effect of ions, which has not been comprehensively studied in previous ICEO simulations \cite{feng2020recent}. Previous analyses of the steric effects have been limited to theories \cite{kilic2007steric,storey2008steric,bazant2009towards}; those theories either neglect the interactions from the walls and result in large values of ion sizes when fitting to experimental data \cite{storey2008steric}, or make simplified assumptions on the fitting parameters to facilitate better fitting \cite{bazant2009towards}. 

To illustrate the importance of steric effects in ICEO simulations, in Figure \ref{fig:ux_reverse}(a,b) we present the flow field and charge distribution from the PNP simulation for Case~C at 0.15~M.
The images show the development of secondary vortices just above the bottom boundary, appearing as small ``bumps" on the streamlines and magnified in Fig.~\ref{fig:ux_reverse}(b), which are not present in the DISCOS simulations shown in Fig.~\ref{fig:flow}. 
These secondary vortices in the PNP simulation lead to a strong localized flow reversal in the horizontal velocity profile right above the bottom boundary (Case~C at 0.15~M) as shown in Figure \ref{fig:ux_reverse}(c) (dashed line). On the contrary, the DISCOS simulation at the same condition (Figure \ref{fig:ux_reverse}(b), symbols) shows no such flow reversal within the margin of statistical error.

We conjecture that these secondary vortices arise from the lack of steric effects in the PNP model.  As discussed in Sec.~\ref{subsec:scaling}, for a continuum algorithm like the PNP method, there is a high ion concentration near the boundary that interacts with electric field singularity at the transition of boundary condition from dielectric to metal, resulting in the formation of nonphysical vortices.
With the DISCOS method, steric effect keeps ions away from it, giving more realistic behaviors. This velocity reversal near the singularity point may also be the reason behind the velocity scaling turnover in the PNP simulation, as shown in Fig.~\ref{fig:ux_reverse}(d), where the ratio between the reversed horizontal velocity (defined as the maximum velocity in the reverse direction) and the overall peak velocity in the channel (depicted on the blue dashed line in Fig.~\ref{fig:scaling}) is plotted against the electric fields. The reversed velocity stays around 16\% of the peak velocity for $E_x^{\textrm{ext}}\leq 7.5\times 10^4$~V/cm, which corresponds to the quadratic velocity scaling regime in Fig.~\ref{fig:scaling}, suggesting that the steric effect plays a less prominent role. This ratio starts to rise for higher field strengths; the electric field beyond which this ratio rises matches quantitatively with the transition of the velocity scaling in Fig.~\ref{fig:scaling}. The magnitude of the reversed velocity becomes comparable to the peak velocity at $E_x^{\textrm{ext}}= 1.5\times 10^6$~V/cm; the velocity reversal becomes so dominant beyond this point that the PNP simulation becomes numerically unstable. Moreover, the steric effect can also potentially explain the suppressed velocity scaling observed in DISCOS simulations. As mentioned above, the DISCOS simulations show the development of a secondary layer of co-ions from overcharging, as shown in Fig.~\ref{fig:peak_vel_rhoE}-\ref{fig:1e14}. The forces on these co-ions act in the direction opposite to the circulation in the primary vortex pair, which could potentially account for the reduced velocity scaling.

\section{Conclusions}\label{sec:conclusion}
In this article we present simulation results of ICEO using a newly developed, mesoscale fluid model called DISCOS. In earlier work this method was shown to generate reliable results for equilibrium ionic structures and regular electro-osmotic flows in a nanoscale channel. Here we have extended to highly non-linear flows of ICEO by implementing a boundary condition that switches between Dirichlet and Neumann conditions for the electrostatics to model a metal strip sandwiched by a dielectric substrate at the bottom of the channel. Our simulations show qualitatively correct results, namely, that a counter-rotating vortex pair is formed for all electric field strengths. A comparison to the deterministic PNP method demonstrates a fundamental advantage of DISCOS over traditional continuum methods, in that DISCOS correctly captures the steric effect, which is a crucial mesoscale phenomena that leads to unique features in the flow field and charge distribution. DISCOS also naturally avoids the charge density and electric potential singularity near the edges of the metal strip, allowing it to simulate much larger electric fields than the deterministic PNP method. 

By examining a wide range of electric field strengths, we found that the characteristic velocity in our simple ICEO setup transitions from a quadratic scaling of velocity with electric field to an almost linear scaling as the induced $\zeta$-potential exceeds the thermal voltage. Simulating a high molarity system shows a similar trend. Comparison between the DISCOS and the PNP methods shows that although these methods show a qualitatively similar trend in velocity scaling, they predict solution structures that are quite different. 
For example, the location of the peak charge density is distinctly different between DISCOS, for which the peak shifts toward the center-line with increasing field strength, and PNP methods where the peak always occurs at the edge of the metal plate. Furthermore, a layer of co-ions to the surface charge forms in DISCOS method but not in PNP method.

Although both methods show a transition to linear scaling for high field strengths, a detailed examination of the transition suggests that the physical mechanisms are different.
We explored several mechanisms from the literature that are potentially responsible for the velocity scaling transition. We demonstrated that charge-induced thickening plays a minor role in the transition of velocity scaling, as the viscosity near the location of maximum velocity is essentially unchanged from the bulk solvent viscosity. The $\zeta$-potential does not explain the transition either, as it shows linear dependence on the electric field before and after the transition. 
The key factor in the transition mechanisms for the two methods is steric effects.
In the PNP method, the lack of steric effects allows ions to interact with the singularity in the electric field, creating a secondary vortex pair that suppresses the velocity scaling. In the DISCOS method, explicitly modeling steric effect leads to a weak interaction of ions with the singularity.  It also lead to an overcharging effect manifested in the accumulation of co-ions that can also suppress the characteristic velocity. 

One potential avenue for future investigation of alternating current electro-osmosis (ACEO) to elucidate the origin of the velocity scaling turnover, which is typically conducted in experiments. However, because of the charging-discharging process of ACEO, the average charge density within the double layer is smaller and the noise is higher, resulting in longer simulations to get better statistics. Therefore we are developing novel numerical algorithms to improve computational efficiency. One direction we are currently focusing on is construction of a hybrid DISCOS-FHD algorithm where DISCOS is utilized near the double layer while FHD is utilized in the bulk flow. 

\section{Acknowledgements}\label{sec:acknowledgement}
This work was supported by the U.S.~Department of Energy, Office of Science, Office of Advanced Scientific Computing Research, Applied Mathematics Program under contract No.~DE-AC02-05CH11231. 
This research used resources of the National Energy Research Scientific Computing Center, a DOE Office of Science User Facility supported by the Office of Science of the U.S. Department of Energy under Contract No.~DE-AC02-05CH11231.




\clearpage

\begin{appendix}

\section{Appendix: Numerical Accuracy and Wet Percentage}\label{appx:dry}
To save computational cost we establish the optimal wet percentage for our simulations. Wet percentage essentially describes an effective hydrodynamic radius, which is proportional to the mesh size. By definition, the 100\% wet case is the most accurate, correctly capturing the hydrodynamic radius of ions. Increasing the mesh size (coarser mesh) means reducing wet percentage, thus overestimating the hydrodynamic radius, so in DISCOS we introduce a correction (recall this is the dry diffusion contribution) to recover the desired total diffusion. 

\begin{figure}[h!]
    \centering
    \includegraphics[width=0.5\textwidth]{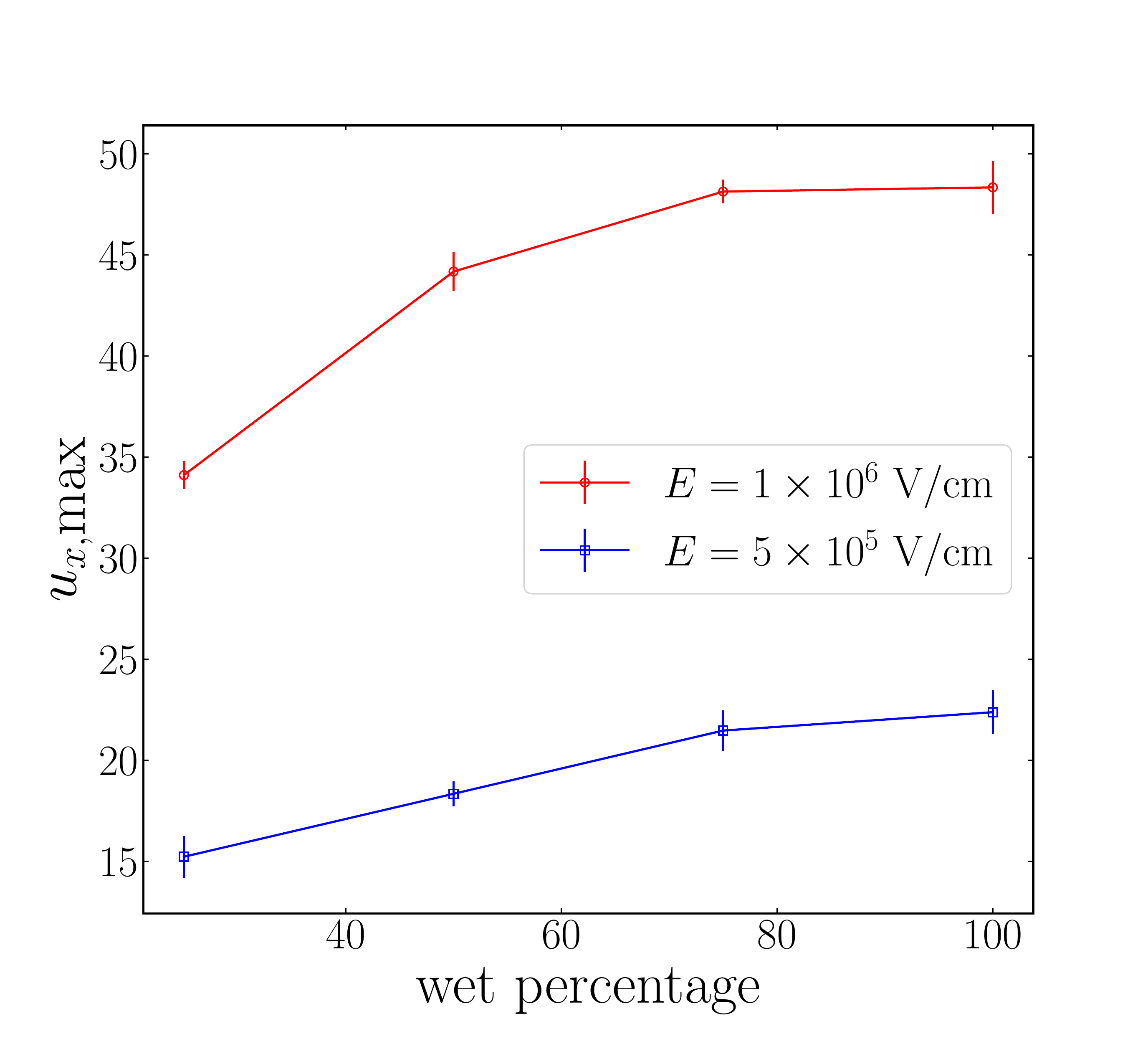}
    \caption{Comparison of the effect of wet percentage on the measured peak velocity at two different electric field. Error bars represent one standard error.}
    \label{fig:wet}
\end{figure}

Figure~\ref{fig:wet} shows a plot of peak velocity versus wet percentage for two different magnitudes of the applied $E$-field (Cases A and B). It is observed that the 25\%-wet and 50\%-wet cases significantly underestimate the peak velocity, suggesting that substantial contributions from many-body hydrodynamic interactions are missing and dry diffusion alone cannot recover them. However, beyond 75\%-wet the peak velocities are converged to the 100\%-wet case, within statistical error. This indicates that the grid of a 75\%-wet system is sufficient to capture the most of the many-body hydrodynamic interactions, so we use that system to save computational cost while maintaining the physical accuracy. As such, all the the results presented in this paper are from simulations that are 75\%-wet.

\end{appendix}

%

\end{document}